\documentstyle[12pt]{article}
\newcommand{\sect}[1]{\setcounter{equation}{0}\section{#1}}
\renewcommand{\theequation}{\thesection.\arabic{equation}}
\begin{document}

\begin{flushright}
{LPTENS-94/35}\\
{NSF-ITP-94/135}\\
{UT-KOMABA-94/21}\\
{\sl December 1994}\\
\end{flushright}
\vskip 2mm
\begin{center}
{\large\bf  UNIVERSAL CORRELATIONS FOR DETERMINISTIC
PLUS RANDOM HAMILTONIANS}
\end{center}
\vskip 3mm
\begin{center}
{{ E. Br\'ezin$^{a)}$, S. Hikami$^{b)}$
and A. Zee$^{c)}$ }\\
\vskip 2mm
{$^{a)}$ Laboratoire de Physique
Th\'eorique{\footnote{
Unit\'e propre du Centre National de la Recherche
Scientifique,
associ\'ee \`a l'Ecole\\
Normale Sup\'erieure et \`a
l'Universit\'e de Paris-Sud}}}\\
{Ecole Normale Sup\'erieure}\\
{24 rue
Lhomond 75231, Paris Cedex 05, France}\\
\vskip 2mm
{$^{b)}$
Department of Pure and Applied Sciences}\\
{ University of
Tokyo}\\
{Meguro-ku, Komaba, 3-8-1}\\
{Tokyo 153, Japan}\\
\vskip 2mm
{$^{c)}$ Institute for Theoretical Physics}\\
{University of California}\\
{Santa Barbara, CA 93106, USA}}
\end{center}
\vskip 5mm
\begin{abstract}

We consider the (smoothed) average correlation between the density of  energy
levels of a disordered system, in which the Hamiltonian is equal to  the sum of
a deterministic $H_{0}$ and of a random potential $\varphi$. Remarkably, this
correlation function  may be explicitly determined in the limit of  large
matrices, for any unperturbed $H_0$ and for a class of probability distribution
$P(\varphi)$ of the random potential. We find a compact representation of the
correlation function. From this representation one obtains readily
the short distance behavior, which  has been conjectured in  various contexts
to be universal. Indeed we find that it is totally  independent of both $H_0$
and $P(\varphi)$.

\end{abstract}
\newpage

\sect{Introduction}

   We consider a Hamiltonian
$ H = H_0 + \varphi$ with a deterministic $H_0$ and a random
$\varphi$,
 in a finite dimensional space in which $H_0$ and $\varphi$
are
 Hermitian $N \times N$ matrices; we are interested in the large N
limit.
The average number of eigenvalues in an interval is known to depend
sensitively upon the spectrum of $H_0$ and upon the probability
distribution
$P(\varphi)$ of the disordered potential$^{1)}$.
 The one-point Green's function and the two-point connected
Green's function are given by
\begin{equation}\label{gz}
   G(z) = < {1\over{N}} {\rm Tr} {1\over{ z - H }} >
\end{equation}
\begin{eqnarray}\label{g2c}
  G_{2c}(z,w) &=&  < ({1\over N} {\rm Tr}{1\over{z - H}}
) ({1\over N}{\rm Tr}{1\over{w - H}} ) >\nonumber\\
&-&  < {1\over N}{\rm Tr}{1\over{ z - H }} >
< {1\over N}{\rm Tr} {1\over{w - H}} >
\end{eqnarray}
We obtain  the correlation function between two eigenvalues from
$G_{2c}(z,w)$
\begin{eqnarray}\label{rho2c}
\rho_{2c}(\lambda,\mu) &=& -{1\over{4 \pi^{2}}}
[G_{2c}(\lambda+i\epsilon, \mu+i\epsilon)+
G_{2c}(\lambda-i\epsilon,\mu-i\epsilon)\nonumber\\
&-& G_{2c}(\lambda+i\epsilon, \mu-i\epsilon)
- G_{2c}(\lambda-i\epsilon,\mu+i\epsilon)]
\end{eqnarray}

In the simpler case in which $H_0$ is zero, it was proved in
earlier papers$^{2),3)}$ that the resulting expression for the connected
two-point Green's function, is independent, up to a scale, of the
probability distribution of the random potential $\varphi$ for arbitrary
values of $\lambda$ and $\mu$.
One word of clarification is needed at this stage. While in ref[2] we
computed $\rho_{2c}(\lambda,\mu)$ directly, in ref[3] we computed
$G_{2c}(z_{1},z_{2})$ by letting N go to infinity first,
with complex $z_{1}$ and $z_{2}$. To obtain the correlation
function we then let the imaginary parts of $z_{1}$ and $z_{2}$ go to
zero. By this procedure we obtain a correlation function
$\rho_{2c}(\lambda,\mu)$ in which a smoothing of the eigenvalues
in a range much larger than $1/N$, but small compared to the average
point spacing, has been performed. This smoothing eliminates all
the non-universal oscillations which are present in $\rho_{2c}(\lambda,
\mu)$ if we had let the imaginary parts go to zero first$^{2)}$ before
letting $N$ go to infinity.

   For given non-zero $H_0$, we could first question whether or not
 $\rho_{2c}$ depends upon the probability distribution of
$\varphi$ and next, how the result depends upon the spectrum of $H_0$.
We shall answer these questions in several steps:
first return for definiteness to the pure random case in which $H_0=0$,
which was studied earlier$^{2)}$.
Then we consider $H_0$ non-zero and a Gaussian $P(\varphi)$, on the
basis of earlier work on the subject.$^{4)}$ We consider next the non-
Gaussian
case with arbitrary $H_0$ and derive first a
closed expression for the one point Green's function. We then apply this
closed expression for two specific examples in which
the eigenvalues of $H_0$ are restricted to $\pm1$ and in which
 $H_0$ has  uniformly spaced levels between $-1$ and $+1$.
Further we derive the two-point connected Green's function
$G_{2c}(z,w)$.
We find that the short distance behaviour is universal.

The universality of the short distance correlation is perhaps expected from
the following heuristic argument. Imagine
switching on adiabatically the potential $\varphi$. Starting from
some initial eigenstate of $H_0$, the potential will induce a succession
of transitions between those eigenstates. For finite time it is expected
that this process depends sensitively on the nature of the spectrum of
$H_0$.
However for large times the system will explore all the eigenstates of
$H_0$ and
one might wonder, or may be even suspect, whether the limit is insensitive
to
$H_0$. Therefore it is natural to consider the limit in which the
eigenvalues
 $\lambda$ and $\mu$ are near each other, and we prove that in this limit
$\rho_{2c}$ is universal.
\sect{ The simple Wigner case}

   We return to the simplest case in which $H_0$ is zero and $P(\varphi)$
is
a Gaussian distribution characterized by
\begin{eqnarray}\label{phi}
     < \varphi_{ij} > & =& 0 \nonumber \\
     < \varphi_{ij}\varphi_{kl} >& =& {v^{2}\over N} \delta_{jk}\delta_{il}
\end{eqnarray}
Much is known about this Gaussian case of course$^{5),6)}$. We have found that
the
correlation function in the large N limit$^{2)}$ may be written in the compact
form
\begin{equation}\label{N2G2c}
N^{2} G_{2c}(z_{1},z_{2}) = - {\partial^{2}\over{\partial z_{1}
\partial z_{2}}}{\rm Log}[ 1 - G(z_{1})G(z_{2})]
\end{equation}
It is convenient for later generalization to define
\begin{equation}\label{uz}
      u(z) = z - G(z)
\end{equation}
given explicitly in this simple case by 
\begin{equation}\label{uz2}
u(z) = {1\over{2}}[ z + \sqrt {z^{2} -
4}]
\end{equation}
Since
$G(z)=1/[z-G(z)]$, (\ref{N2G2c}) is expressed also by 

\begin{equation}\label{N2G2c2}
    N^{2}G_{2c}(z_{1},z_{2}) = {\partial^{2}\over{\partial z_{1}
\partial z_{2} }}{\rm Log}[{u(z_{1}) - u(z_{2})\over{z_{1} - z_{2}}}]
\end{equation}
It is useful to introduce also the
polarization function
\begin{equation}\label{Pi0}
\Pi_{0}(z_{1},z_{2}) = G(z_1)G(z_2)
\end{equation}
We are working here in units in
which $v$, the width of the Gaussian distribution, is one; it is easy
by dimensional analysis to set it back to
any desired value.

There are several ways of deriving (\ref{N2G2c}). In ref[2] it was obtained
through a study of the
asymptotic form of the relevant orthogonal polynomials of high order. In
ref[4] it was
shown that it can be recovered simply by summing the diagrams obtained by
expanding in powers of $v$: they consist of planar connected diagrams made of
ladders and crossed ladders, rainbow-like vertex corrections and full planar
propagators.

 The form that we give
here is not the one which
appears in ref[2], but it is easy to carry out the
differentiations
and verify that it is same. By differentiating (\ref{N2G2c}),
we have
\begin{equation}\label{N2G2c3}
  N^{2} G_{2c}(z_{1},z_{2}) = {1
\over{2(z_{1} - z_{2})^{2}}}
 ( {{z_{1}z_{2} - 4}\over{[(z_{1}^{2} -
4)(z_{2}^{2} - 4)]^{1/2}}}
- 1  )
\end{equation}
from which, using
(\ref{rho2c}), we obtain$^{2),7)}$
\begin{equation}\label{rho2c2}
  \rho_{2c}(\lambda,
\mu) = -{1\over{2 N^{2} \pi^{2}}}{1\over{(\lambda
 - \mu)^{2}}}{ {4 -
\lambda\mu}\over{[( 4 - \lambda^{2})( 4 - \mu^{2}
)]^{1/2}}}
\end{equation}
For $\lambda$ close to $\mu$ it has a singular
behaviour
\begin{equation}\label{rho2c3}
\rho_{2c}(\lambda,\mu) \simeq -{ 1
\over{ 2 \pi^{2}N^{2} (\lambda
- \mu)^{2}}}
\end{equation}
It is
instructive to compare it with the exact expression$^{2)}$, i.e.
without
smoothing. In the limit $\lambda - \mu$ goes to zero with 
\begin{equation}\label{x}
 x = 2\pi N (\lambda - \mu )
\rho({1\over{2}}(\lambda + \mu ))
\end{equation}
fixed.
In this limit the exact
correlation function
becomes$^{6)}$
\begin{equation}\label{rho2c4}
\rho_{2c}(\lambda, \mu)
\simeq -
{1\over{\pi^{2}N^{2}}}{\sin^{2} x
\over{(\lambda -
\mu)^{2}}}
\end{equation}
The smoothing has thus replaced $\sin^{2}x$
by 1/2, thereby producing a double pole.


\sect{General probability distribution without H$_{0}$ }

We now go to
an arbitrary probability distribution for the
disorder.
The
density of eigenvalues is no longer semi-circular; we limit ourselves
here
to spectra which extend over a single segment of length $4a$ on the
real
axis. We choose the origin at the center of this segment. This is automatic
when the
probability distribution $P(\varphi)$ is even;
otherwise
one would have to shift the eigenvalues.
Then we know from
earlier work$^{2),7),8)}$ that, rather remarkably, (\ref{N2G2c2}) is still
valid
with
\begin{equation}\label{uz3}
     u(z) = {1\over{2}}[ z + \sqrt{z^{2} -
4a^{2}} ]
\end{equation}
and hence
\begin{equation}\label{rho2c5}
\rho_{2c}(\lambda,\mu) = - {1\over{2N^{2}\pi^{2}}}
{1\over{(\lambda-
\mu)^{2}}}
 {{4a^{2} - \lambda\mu}\over{[(4a^{2} -
\lambda^{2})(4a^{2} - \mu^{2})]^{1/2}}}
\end{equation}
The non-Gaussian terms of $P(\varphi)$ have stretched the support of the
spectral
density from [-2, 2] to [-2a, 2a], and also deformed the density of eigenvalues
from Wigner's semi-circle
to a polynomial multiplying a square-root$^{1)}$. However, the only
effect of these non-Gaussian terms on $G_{2c}(z_{1},z_{2})$ and
$\rho_{2c}(\lambda,\mu)$ is a simple
rescaling $ z_{1,2} \rightarrow
{z_{1,2}\over{a}}$, $\lambda
\rightarrow {\lambda\over{a}}$, $\mu
\rightarrow {\mu\over{a}}$, followed by the multiplying $G_{2c}$ and
$\rho_{2c}$ by
$(1/a)^{2}$.

  To
derive these results we had to proceed in two steps. First$^{2)}$,
we considered a probability distribution
\begin{equation}\label{Pphi}
P(\varphi) = {1\over{Z}} \exp [ - N {\rm Tr} V(\varphi)
]
\end{equation}
in which $V(\varphi)$ was an arbitrary polynomial of $\varphi$. In
these cases we could  diagonalize $\varphi$,
integrate out the unitary group, and handle the eigenvalues of $\varphi$ by
orthogonal polynomial techniques$^{6),10)}$ (or
Dyson's 2D electrostatic approach$^{9),11)}$).
It was then extended
to
non-unitary invariant measures, or equivalently to a measure in which
${\rm Tr} V(\varphi) $ was replaced by a sum of arbitrary products of
traces of
powers of $\varphi$$^{3)}$.

  If we were to try again to calculate $G_{2c}$ by expanding the
non-Gaussian terms of the measure and summing the planar diagrams,
we would have to deal with expectation values
of the type $<{\rm
Tr}\varphi^{n}{\rm Tr}\varphi^{m}>$ expanded in powers of the
non-Gaussian
terms.
An equivalent way of stating the results above is the following: Call
$\alpha_{n,m}$ the Gaussian connected expectation
value
\begin{equation}\label{alphanm}
    \alpha_{n,m} = < {\rm Tr} \varphi^{n}
{\rm Tr }\varphi^{m} >_{c,0}.
\end{equation}
For arbitrary n and m it is
rather cumbersome to determine
these numbers. They are of order one in
the large $N$ limit, whereas the
disconnected part is of order $N^{2}$. If
needed explicitly, we could obtain
them by expanding their generating
function
\begin{equation}\label{N2G2c4}
   N^{2} G_{2c}(z_{1},z_{2}) =
\sum_{n,m} \alpha_{n,m} z_{1}^{-n-1} z_{2}^
{-m-
1}
\end{equation}
given by (\ref{N2G2c2}), in inverse powers of the z's.

We now
consider the same expectation value $<{\rm Tr} \varphi^{n}{\rm Tr}\varphi^{m}
>_{c}$
for the non-Gaussian case. The result of (3.2) is equivalent to the
following
statement
\begin{equation}\label{Trphi}
      <{\rm Tr} \varphi^{n}
{\rm Tr} \varphi^{m} >_{c} = {a}^{n+m}\alpha_{n,m}
\end{equation}
in
which $2a$ is again the end point of the support of the spectral density.

A direct derivation of these identities is non-trivial. For specific values of
$m$ and $n$
they may be extracted from what is known about $G(z)$. See Appendix A.

\sect{One-point Green's function with non-zero H$_0$}

    We now have a
given unperturbed Hamiltonian $H_0$, with eigenvalues
$\epsilon_{i}$, $i
= 1, ...,N$, with $N$ large as always. The one-point Green's function, or the
average
resolvent, was
first found by Pastur$^{12)}$ for the Gaussian distribution
$P(\varphi)$.
 Define the deterministic
unperturbed one-point Green's
function
\begin{eqnarray}\label{G0}
   G_0(z) &=&
\frac{1}{N} {\rm Tr} \frac{1}{z - H_0} \nonumber\\
&=& \frac{1}{N} \sum_{i} \frac{1}{z - \epsilon_{i}}
\end{eqnarray}
Then the full one-point Green's function
\begin{eqnarray}\label{gz2}
   G(z) &=&  <
{1\over{N}} {\rm Tr} \frac{1}{z - H}  >\nonumber\\
\end{eqnarray}
is obtained by solving the implicit equation
\begin{equation}\label{gz3}
G(z) = G_0(z - G(z))
\end{equation}
In other words, the  one-point Green's function is determined by
\begin{eqnarray}\label{gz4}
   G(z) &=&  <
{1\over{N}} {\rm Tr} {1\over{z - H_0 - \varphi}} >\nonumber\\
&=& {1\over{N}} \sum_{i} {1\over{z - \epsilon_{i} -
G(z)}}
\end{eqnarray}

    In the simplest case, in which $H_0 = 0$, the
above equation becomes
a simple quadratic equation for G, from which
one recovers immediately
Wigner's semi-circle law for the imaginary part
of G, which is proportional to
the density of eigenvalues. It is easy to give a
diagrammatic
proof$^{4)}$ of (\ref{gz4}). In the large $N$ limit, the planar
diagrams are
the rainbow diagrams whose sum leads directly to (\ref{gz4}).

The
self-energy part $\Sigma (z)$ of the one-point Green's function $G(z)$
is simply given for a Gaussian distribution $P(\varphi)$
by
\begin{equation}\label{Sigmaz}
   \Sigma(z) = G(z)
\end{equation}
For
the non-Gaussian case, the self-energy part $\Sigma(z)$, defined by
\begin{equation}\label{gz5}
  G(z) =
{1\over{N}}\sum_{i}{1\over{z - \epsilon_{i} -
\Sigma(z)}}.
\end{equation}
 is of course no longer simply equal to $G(z)$.

Here we will restrict ourselves to the non-Gaussian probability distribution
for
\begin{equation}\label{Pphi2}
     P(\varphi) =
{1\over{Z}}\exp [ -N( {1\over{2}}{\rm Tr} \varphi^{2} + {g }
{\rm Tr}
\varphi^{4} )]
\end{equation}
To treat the non-Gaussian term $g{\rm Tr}
\varphi^{4}$, it is convenient
to consider an
``equation of motion''
obtained by shifting the random matrix by an arbitrary matrix
$\varphi$ to $\varphi
+ \epsilon$ in the integral which gives $<(1/(z - H_{0} - \varphi))_{ij}>$.
The invariance of this integral under this shift tells us that to first order
in $\epsilon$, the coefficient of $\epsilon_{nm}$ must vanish:
\begin{eqnarray}\label{epsilon}
&& < ( {1\over{z - H_0 -
\varphi}} )_{in}
 ({1\over{z - H_0 -\varphi}} )_{mj} > -\nonumber\\
&& N < ({1\over{z - H_0 -\varphi}}
 )_{ij} (\varphi_{mn} + 4 g
(\varphi^{3})_{mn} ) > = 0
\end{eqnarray}
Setting $n = i$ and $m = j$, and summing over these indices, we
obtain  the following equation with $H =
H_{0} + \varphi$,
\begin{equation}\label{TrzH}
< ({1\over{N}} {\rm Tr}
{1\over{ z - H }})^{2} >
= < {1\over{N}} {\rm Tr} [ {1\over{ z - H}}  (
\varphi + 4 g \varphi^{3}  )  ]  >
\end{equation}
By the definition of two-point connected Green's function, we write above
equation as
\begin{equation}\label{NG2}
N G^{2}(z) + N G_{2c}(z,z) =  < {\rm
Tr} ({1\over{z - H}} \varphi ) >
+ 4 g  < {\rm Tr} ({1
\over{z - H}}
\varphi^{3} )  >
\end{equation}
where $G_{2c}(z,z)$ is order of $1/N^{2}$.

The first term on the right hand side of this equation may be rearranged by
writing
$\varphi
= (\varphi+H_0-z)+(z-H_0)$:
\begin{eqnarray}\label{TrzH2}
{1\over{N}} <
{\rm Tr} ({1\over{z - H_0 -\varphi}}\varphi ) > &=& -1 +
{1\over{N}}  <{\rm Tr}{{1}\over{z - H_0 - \varphi}}
{ (z - H_{0} )} >\nonumber\\
&=& -1 + {1\over{N}}\sum_{i}{{z - \epsilon_{i}}\over{z
- \epsilon_{i}
- \Sigma(z)}}\nonumber\\
&=&
\Sigma(z)G(z)
\end{eqnarray}
with the self-energy $\Sigma(z)$ defined above.  Repeating the same procedure,
we write the second term on the right hand side of (4.6)
\begin{eqnarray}\label{TrzH3}
 < {\rm Tr}
 ({1\over{z - H_0 - \varphi}} \varphi^{3} ) > &=& - \left <
{\rm Tr} \varphi^{2} \right >\nonumber\\
&& \!\!\!\!\!\! \!\!\!\!\!\!\!\!\!\!\!\!\!\!\!\!\!
 + < {\rm Tr} [ {1\over{z -
H_{0} - \varphi}}(z - H_0)\,\varphi(z - H_0) ]  >
\end{eqnarray}
We have thrown away the
term $<{\rm Tr}\varphi(z - H_0)>$ since it is zero by being odd in $\varphi$.

We will now consider the relevant Feynman diagrams.
As in ref[4] we find it convenient to use the language of
large $N$ QCD to describe the diagrams.
Let us introduce the  the gluon-quark-quark vertex
function $\Gamma(z)$ defined to include gluon self-energy
corrections. In other hands, it is one-particle-irreducible
in the quark lines, but not in the gluon line.
Then we can write the second term in the equation above as
\begin{eqnarray}\label{Sigmaz2}
& &\sum_{i}\sum_{j}(z-\epsilon_{i})(z-\epsilon_{j})
 < ({1\over {z-H_0 -\varphi}} )_{ij}\varphi_{ji} >
\nonumber\\
&  =& {1\over{N}}
\sum_{i}\sum_{j}
(z-\epsilon_{i})(z- \epsilon_{j}){1\over
{(z - \epsilon_{i} - \Sigma(z))  (z - \epsilon_{j} - \Sigma(z)) }}\Gamma(z)
\nonumber\\
&  =&N \Gamma(z) (1+ \Sigma(z) G(z))^2
\end{eqnarray}

As shown in Fig. 1, in the planar limit the vertex
function $\Gamma(z)$, the (quark) self-energy $\Sigma(z)$,
and the Green's function or quark propagator $G(z)$, are related by
\begin{equation}\label{Sigmaz3}
\Sigma(z) = \Gamma(z) G(z)
\end{equation} Using this fact, and putting everything together, we find that
$G(z)$,  for arbitrary
non-zero
$H_0$ and for a non-Gaussian probability distribution for $\varphi$,  is
determined by
\begin{equation}\label{G2z}
G^{2}(z) =
\Sigma(z) G(z) - 4 g x + 4 g{\Sigma(z)\over{G(z)}}(1+\Sigma(z)G(z)
)^{2}
\end{equation}
where we denote $<{\rm Tr} \varphi^{2}>$ by $x$.
This quantity $x$ does not depend on
$H_0$ and is simply
related to the end point $a(g)$ of the spectrum for the $H_0 = 0$
case by
\begin{equation}\label{x2}
       x = {a^{2}\over{3}} ( 4 - a^{2} )
\end{equation}
where $a(g)$ satisfies the following
equation$^{1)}$
\begin{equation}\label{ag}
   12 g a^{4} + a^{2} = 1
\end{equation}

    Thus we have accomplished our goal of finding the equation
(\ref{G2z})
which determines  the one-point
Green's function $G(z)$ in general. Note that the self-energy $\Sigma$ which
appears in (\ref{G2z}) is itself related to $G(z)$ as described in (\ref{gz5}).
These two coupled equations are not explicitly solvable for an arbitrary $H_0$.

We now consider two  simple
cases
as examples, specified by the eigenvalues {$\epsilon_i$} of
$H_0$: i)   $\epsilon_{i} = \epsilon$ for $i = 1,...,N/2$,
and $\epsilon_{i} =
-\epsilon$
for $i=(N/2)+1,...,N$, with $N$ even.
ii) the $\epsilon_{i}$ are
uniformly spaced over an interval from $-\alpha$ to $\alpha$.

 For the first case i), we have from (\ref{gz5}),
\begin{equation}\label{G1}
  G = {1\over{2}} ( {1\over{z - \epsilon - \Sigma}}
+ {1\over{z + \epsilon - \Sigma}} )
\end{equation}
In the case $g = 0$, (\ref{G2z}) tells us simply
$\Sigma = G$ and thus we obtain the cubic equation
\begin{equation}\label{cubic}
    ( z - G
)^{2} G + ( 1 - \epsilon^{2}) G - z = 0
\end{equation}
The imaginary part of $G(z)$ is
represented in Fig. 2 by a solid line for the particular case $\epsilon = 1$.
The end-points of the spectrum occur at  $z =
\pm\sqrt{27}/2$. The point $ z = 0$ becomes
also the end point for $g = 0$. For
$ g \neq
0$, we consider the following representative values for $\epsilon = 1$,
a) $g = 1/2$, $x$ = 11/27,
b) $g = -1/48$, $x = 4/3$, and solve
numerically. We
represent the
imaginary part of $G(z)$ in
Fig. 2 by a dotted line and by a broken line, respectively.
We have a fifth
order polynomial equation for $\Sigma$ by substituting
(\ref{G1}) to (\ref{G2z}).
At the critical value $g = -1/48$,
we find the imaginary part of
$\Sigma$ and $G$,
vanish at $z_{c} \cong 3.225$  with the exponent
3/2,
\begin{equation}\label{ImG}
    {\rm Im} G(z) \simeq ( z_{c} -
z)^{3/2}
\end{equation}
which shows the same singularity as the $H_{0}=0$
case at the edge.$^{1),14)}$

In the second example ii), we have from (\ref{gz5}),
\begin{equation}\label{z1}
z = \Sigma(z) + \alpha \coth (\alpha G(z))
\end{equation}
We consider further the case $\alpha = 1$.The
imaginary part of G(z)
is evaluated numerically, and represented in
Fig. 3.
At the critical value $g = -1/48$, the imaginary part of $G(z)$
shows a same singularity at the edge as before. Note that the
value $g = -1/48$ always remains critical, independent of $H_{0}$.
This is due to the singular behavior of $x$ given by (4.16) near
$g_{c} = -1/48$, which is independent of $H_{0}$.
 We have considered the non-Gaussian probability distribution
given by (\ref{Pphi2}). It is easy to apply the present method to
other non-Gaussian $P(\varphi)$ case.

\sect{ Two-point correlation function with non-zero H$_0$ and  $\varphi$
Gaussian}

We now finally come to the two-point connected Green's function, or
correlation function, $G_{2c}(z,w)$. In an  earlier work$^{4)}$ we have shown
that for a Gaussian distribution $P(\varphi)$ we can determine $G_{2c}(z,w)$
as follows. First, we define 
\begin{equation}\label{gi}
g_{i}(z) = {1\over{z - \epsilon_{i} -  \Sigma(z)}}
\end{equation}
For the Gaussian case, the self-energy $\Sigma = G(z)$. For notational
convenience we also define the ``circle-product''
\begin{equation}\label{circle}
G^{n}(z) \circ G^{m}(w) =  {1\over{N}}\sum_{i=1}^{N}
g_{i}^{n}(z)g_{i}^{m}(w)
\end{equation}

Then in the large N  limit, we found 
\begin{eqnarray}\label{N2G2c5}
N^{2} G_{2c}(z,w) &=& ( {G^{2}(z)\circ G^{2}(w)\over{1 - G(z)\circ
G(w)}}\nonumber\\
&+& {(G^{2}(z)\circ  G(w))(G(z) \circ G^{2}(w))\over{[ 1 -  G(z) \circ G(w)
]^{2}}})  \nonumber\\ &\times& {1\over{1 - G(z)\circ G(z)}}{1\over{1 -
G(w)\circ G(w)}}
\end{eqnarray}
This result was derived by  summing all
the relevant planar diagrams, consisting of generalized ladders,  with
arbitrary
cyclic permutations of the rungs, with fully dressed planar propagators and
planar, i.  e. rainbow-like, vertex corrections. (We have  again set  $v$, the
width of the random distribution, to one).

Surprisingly enough, we can show that this rather complicated expression
can again be written simply in the form
\begin{eqnarray}\label{N2G2c6}
    N^{2} G_{2c}(z_{1},z_{2}) &=&
-{\partial^{2}\over {\partial z_{1} \partial z_{2}}}{\rm Log}
 ( 1 -
G(z_{1})\circ G(z_{2}) )
\end{eqnarray}
Note that this expression has the same form as (\ref{N2G2c}) but with the
ordinary product replaced by the circle product. Defining $u(z) = z -G(z)$ as
before, and using the identity
\begin{equation}\label{identity}
     1 - G(z)\circ G(w) = { z -
w \over{ u(z) - u(w) }}
\end{equation}
we can also write
\begin{equation}\label{N2G2c7}
    N^{2} G_{2c}(z_{1},z_{2}) =
{\partial^{2}\over{\partial z_{1}
\partial z_{2}}} {\rm Log}  ( {u(z_{1})
- u(z_{2}) \over
{z_{1} - z_{2} }} )
\end{equation}
just as in (\ref{N2G2c2}).

To derive these compact representations, we simply differentiate
(\ref{N2G2c6}) to obtain the previous result (\ref{N2G2c5}),using the
identity
\begin{equation}\label{identity2}
      {d G(z)\over
{d z}} = - {G(z)\circ G(z)\over{1
     - G(z)\circ
G(z)}}
\end{equation}

\sect{ Two-point correlation function with non-zero H$_0$ and  $\varphi$
non-Gaussian}

The remarkable existence of such a compact representation as (\ref{N2G2c7})
 naturally prompts us to ask whether this representation also holds for a
general non-Gaussian distribution
$P(\varphi)$. In this section, we answer this question partially by studying
$G_{2c}(z, w)$ for
the distribution 
\begin{equation}\label{Pphi3}
P(\varphi) =
{1\over{Z}}\exp [ -N ( {1\over{2}}{\rm Tr} \varphi^{2} + {g }
{\rm Tr}
\varphi^{4}  ) ]
\end{equation}
to first order in $g$.

As contributions of order g, there are six different classes
of diagrams, with their typical representatives shown
in Fig. 4. The contributions of these six classes of
diagrams to $G_{2c}(z,w)$ are
\begin{eqnarray}\label{Da}
    D_{a} &=& - 8
g {\partial^{2}\over{\partial w \partial z}} [{G(z)\circ
          G(w)\over{1
- G(z)\circ G(w)}} ]\nonumber\\
    D_{b} &=& - 4 g
{\partial^{2}\over{\partial w \partial z}}
           [ G(z)G(w)  ]\nonumber\\
D_{c} &=& - 4 g {\partial^{2}\over{\partial w \partial z}}
           [
{G(z)\circ G(w)\over{ 1 - G(z)\circ G(w) }}G(z)G(w) ]\nonumber\\
D_{d} &=& - 4 g {\partial^{2}\over{\partial w \partial z}}
           [
{G(z)\circ G(w)\over{1 - G(z)\circ G(w)}}(G^{2}(z) + G^{2}(w)) ]
\nonumber\\
    D_{e} &=& - 8 g {\partial^{2}\over{\partial w \partial
z}}
          [ {G(z)\over{ 1 - G(z)\circ G(z)}}{G^{2}(z)\circ G(w)\over{
1 - G(z)\circ G(w)}}\nonumber\\
&+&  {G(w)\over{1 - G(w)\circ
G(w)}} {G^{2}(w)\circ G(z)\over{1 - G(z)\circ G(w)}} ]\nonumber\\
    D_{f} &=& - 4 g {\partial^{2}\over{ \partial w
\partial z}}
[ {G^{3}(z)\over{1 - G(z)\circ G(z)}}{G^{2}(z)\circ
G(w)\over{
1 - G(z)\circ G(w)}}\nonumber\\
&+&
{G^{3}(w)\over{ 1 - G(w)\circ G(w)}}
           {G^{2}(w)\circ
G(z)\over{1 - G(z)\circ G(w)}} ]
\end{eqnarray}

To order $g$, the one-point Green's function $G(z)$ that appears here can of
course be taken to be the lowest order Green's function, which we
denote by $G_0(z)$ (note that this $G_{0}(z)$ is different from the one defined
in (4.1), where we used it for the deterministic unperturbed one-point
Green function.)
For the sake of notational simplicity, however, we will omit the subscript
$0$ here. In other words, $G(z)$ is determined by
\begin{equation}\label{Gz}
G(z) = {1\over N} \sum_i {1\over z-\epsilon_i - G(z)}
\end{equation}
Summing these terms
$D_{a}, ..., D_{f}$ and adding this to the
unperturbed term $N^2 G_{2c}(z,w) = -
{\partial^{2}\over{\partial w \partial z}}
               {\rm Log} ( 1 -
G(z)\circ G(w))$ we
obtain
\begin{eqnarray}\label{N2G2c8}
     N^2 G_{2c}(z,w) &=& -
{\partial^{2}\over{\partial w \partial z}}
               {\rm Log} ( 1 -
G(z)\circ G(w)\nonumber\\
        &+&  4 g  [ G(z)\circ G(w) ( 2 + G^{2}(z) + G^{2}(w)
)\nonumber\\
         &+&  G(z) G(w) + G^{2}(z)\circ G(w)  ({2 G(z) +
G^{3}(z)\over
         {1 - G(z)\circ G(z)}} )\nonumber\\
        &+&
G^{2}(w)\circ G(z)  ( {2 G(w) + G^{3}(w)\over { 1 - G(w)
        \circ
G(w)}} ]) + O(g^{2})
\end{eqnarray}
     We note, that to lowest order in $g$,
\begin{eqnarray}
     &&G^{2}(z)\circ G(w) = {{G(z)\circ
G(z) - G(z)\circ G(w)}\over{w - z
            - G(w) + G(z)}}\nonumber\\
&       =& {{G(z)\circ G(z) - 1}\over{w - z - G(w) + G(z)}}
              + {{w -
z}\over{( w -z -G(w) + G(z)
)^{2}}}
\end{eqnarray}
and
\begin{equation}
     G(z)\circ
G(w) = {{G(z) - G(w)}\over{ w - z - G(w) +
G(z)}}
\end{equation}
Then
(\ref{N2G2c8}) may be written as
\begin{eqnarray}
    &&N^{2}  G_{2c}(z,w) =  -
{\partial^{2}\over{\partial w \partial z}}
          {\rm Log} [( {{z -
w}\over{ z - w + G(w) - G(z)}})\nonumber\\
         &     \times& \biggl( 1 + 4 g
G(z) G(w) - 4 g {G(z)(2 + G^{2}(z))\over
        {(z -w - G(z) + G(w))( 1
- G(z)\circ G(z))}}\nonumber\\
       &     +& 4 g {G(w)( 2 +
G^{2}(w))\over{( z -w - G(z) + G(w))( 1 - G(w)
          \circ
G(w))}}\biggr)]
\end{eqnarray}
The last two terms of above equation come from
parts of
diagrams of classes (e) and (f).

We would now like to re-express this result using the exact one-point Green's
function as far as possible.
Using the notation $G_{0}(z)$ for the one-point Green function for
non-zero $H_{0}$ and for the Gaussian distribution ( $g = 0$),
we find the self-energy $\Sigma(z)$ of the non-Gaussian distribution $P(
\varphi)$ is given by
\begin{equation}
\Sigma(z) = G_0(z) - {{8 g G_0(z) +4gG_0^{3}(z)}\over{ 1 - G_0(z)\circ G_0(z)}}
 + O(g^{2})
\end{equation}
Then the last two terms of (6.7) are absorbed by the expression of the
self-energy $\Sigma(w)$ and $\Sigma(z)$.

Therefore, we obtain finally the following simple expression, up to order $g$,
\begin{eqnarray}\label{N2G2c9}
    &&N^2 G_{2c}(z,w) = -
{\partial^{2}\over{\partial z \partial w}}
     {\rm Log} [ ( {{ z - w}\over
{z - w - \Sigma(z) + \Sigma(w)}}
) ( 1 + 4 g G(z)G(w)) ]
\nonumber\\
&      =&
-
{\partial^{2}\over{\partial z \partial w}}
     {\rm Log} [ ( {{ z - w}\over {u(z) - u(w)}}
) ( 1 + 4 g G(z)G(w)) ] +
O(g^{2})
\end{eqnarray}
where we now define
\begin{equation}
     u(z) = z - \Sigma(z)
\end{equation}
as the appropriate generalization of (2.3).

We have thus obtained the same form as in (\ref{N2G2c7})
except for an extra factor
of $( 1 + 4 g G(z)G(w))$. This extra factor, however, does
not contribute to the double pole in
$\rho_{2c}(z,w)$ defined by (1.3). In Appendix B, we show that
this expression is consistent with the known result for $H_{0} = 0$
case. Since this factor appears inside log, we separate this term as
 $ - 4 g (\partial_{z} G(z))(\partial_{w} G(w))$. This term seems
to be unconnected part for two-point Green's function. Thus if we
neglect this factorized term, we have the same form as (2.5).
 The general expression for the connected two point Green's function
may be written as
\begin{eqnarray}
 G_{2c}(z,w) &=& - {\partial \over{\partial z}}{\partial\over{\partial w}}
{\rm Log} [ 1 - G(z)\circ G(w) \Gamma^{(1)}(z,w) ] \nonumber\\
 &+& {\partial\over{\partial z}}{\partial\over{\partial w}}
G(z)G(w)\Gamma^{(2)}(z,w)
\end{eqnarray}
The second term represents the  diagram class like $b$ in Fig. 4.
The fact that the terms in (\ref{Da})
all collect into the relatively simple form of (\ref{N2G2c9})
suggests to us that some variation of this form may
in fact hold to all orders in $g$,
and possibly even to arbitrary $P(\varphi)$.
We do not have a proof of all this
tempting conjecture at this point.


\sect{ Discussion}

Thus far, we have considered the correlation function
$\rho_{2c}$ for arbitrary values
of $\lambda$
and $\mu$, in which case all four terms on the right hand side of  (1.3)
contribute.
In the limit in which $\lambda - \mu$ tends to zero, one gets a singularity
which is
entirely due to the Green's functions with
opposite
signs of their infinitesimal imaginary parts.
Indeed $(u(z) - u(w))/(z - w)$
is not singular when $w$ approaches
$z$, with $z$ and $w$ on the same
side of the cut. However if the
imaginary parts of $z$ and $w$ are
opposite, and when they
approach the real axis on the support of the
spectral density of the
resolvent, the ratio blows up at short distance.
The
residue of the singularity depends upon the imaginary part of $u$,
but it
disappears when we take the derivative of the logarithm.
Therefore if
$\lambda$ and $\mu$ are such that $\rho(\lambda)$ and
$\rho(\mu)$ are
non-zero, when $\mu$ is close to $\lambda$ we have
a double pole with a
universal residue minus one:
\begin{equation}
       N^{2}
G_{2c}(\lambda\pm i\epsilon,\mu\mp i\epsilon) \cong
{-1\over{(\lambda
- \mu)^{2}}}
\end{equation}
Consequently, for arbitrary $H_{0}$, we have as in the
simple Wigner ensemble a
behaviour
\begin{equation}
    \rho_{2c}(\lambda,\mu) \cong -
{1 \over{ 2 \pi^{2} N^{2} (\lambda -
\mu)^{2}}}
\end{equation}
which, remarkably enough, is totally
independent of
any specific
characteristic of the problem. We have proved this for a distribution
$P(\varphi)$ defined with a quartic $V$  to lowest order in the quartic
coupling, but we are tempted to conjecture that this short distance
universality in fact holds to all orders, and perhaps even for arbitrary $V$.

We close with a  few
concluding remarks.

(1) We have checked that the same
result (2.2) holds for a
model that we have considered recently$^{15)}$,
in which random
matrices are made of independent random blocks,
as
when they are attached to a lattice. In that case the result comes
out
immediately from the explicit expression given in that work. We obtain
\begin{eqnarray}
    (CN)^{2} G_{2c}(z_{1},z_{2}) &=&
-{\partial^{2}\over {\partial z_{1} \partial z_{2}}} \sum_k
{\rm Log}
( 1 - \epsilon_k
G(z_{1}) G(z_{2}) )
 \end{eqnarray}
with $C$ the number of lattice sites and $\epsilon_k$
the single particle Bloch energies as defined in ref[15].
Similarly, in ref[4] we have considered
a situation in which the random matrix $\varphi$
depends on a parameter called time. Again, the explicit
expression given in that work may be written in the form
\begin{eqnarray}
    N^{2} G_{2c}(z_{1},z_{2}) &=&
-{\partial^{2}\over {\partial z_{1} \partial z_{2}}}{\rm Log}
( 1 - e^{-u(t)}
G(z_{1})G(z_{2}) )
 \end{eqnarray}
where $u(t)$ is a function of time defined in ref[4].
In fact the universal
form discussed in this paper seems to hold in every problem with some random
features.

 (2) As noted for the simple Wigner case the short distance singularity
 that one finds
in (7.1) for the smoothed correlation function is spurious:
the true
$\rho_{2c}(\lambda,\mu)$ is finite at
short distance, but the smoothing
has replaced some vanishing numerator by
a constant average.  This
justifies the procedure followed in the literature in
computing the universal
fluctuations in mesoscopic systems:
the smoothed function correlation is used with
some cut-off at short distance.
An integration by part$^{16)}$,
particularly easy since $G_{2c}$
takes the form of a second derivative,
allows one to return to
the integrable logarithmic singularity. The short
distance cut-off
may then be removed.

(3) Consequently the interesting
feature of these results is not
so much the universal nature of the short
distance spurious
singularity, but the general representation (2.2) of the
two-point
Green's function.

\newpage


\begin{center}
{\bf Acknowledgement}
\end{center}
We
thank the support by the cooperative research project
between Centre
National de la Recherche Scientifique
(CNRS) and Japan Society for the
Promotion of Science (JSPS).
The work of A. Zee is supported in part by the National Science
Foundation under Grant No. PHY89-04035.

\newpage
\setcounter{equation}{0}
\renewcommand{\theequation}{A.\arabic
{equation}}
{\bf Appendix A: {A calculation of $< {\rm Tr} \varphi^{2n} {\rm
Tr} \varphi^{4} >_{c}$}}
\vskip 5mm

Here we will calculate $< {\rm Tr} \varphi^{2n} {\rm
Tr} \varphi^{4} >_{c}$ for the
probability distribution
\begin{equation}
      P(\varphi) =
{1\over{Z}} \exp [ - N {\rm Tr} ( {1\over{2}}\varphi^{2} + g
\varphi^{4})]
\end{equation} The one-point Green's
function
\begin{equation}
    G(z) = < {1\over{N}} {\rm Tr} (
{1\over{z - \varphi}} ) >
\end{equation}
satisfies the equation$^{13)}$

\begin{equation}
     G^{2}(z) - ( z + 4 g z^{2} ) G(z) + 4 g
z^{2} + {( a^{2} - 2)^{2}\over
     {9 a^{2}}} = 0
\end{equation}
with
the end-points $\pm 2a$ of the spectrum given
by
\begin{equation}
      12 g a^{4} + a^{2} - 1 =
0
\end{equation}

{}From G(z) we obtain, after a tedious but simple
calculation,
\begin{equation}
    < {1\over{N}} {\rm Tr}
\varphi^{2n} > = {(2n)!\over{n! (n+2)!}} a^{2n} [
     2 n + 2 - n a^{2}
]
\end{equation}
Then one has
\begin{eqnarray}
      -{\partial
\over{\partial g}} < {1\over{N}} {\rm Tr} {\varphi^{2n}} >
         &=& <
{\rm Tr} \varphi^{2n} {\rm Tr} \varphi^{4} >_{c}\nonumber\\
         &=&
{(2n)!\over{n!(n+2)!}}n(n+1)( 2 - a^{2}) a^{2 n - 2} {\partial
a^{2}\over{ \partial g}}
\end{eqnarray}
Taking $ {\partial
a^{2}/{\partial g}}  = - { 12 a^{6}/(2 - a^{2})} $
from (3.10), we obtain
as announced
\begin{equation}
      < {\rm Tr} \varphi^{2n} {\rm
Tr} \varphi^{4} >_{c} = a^{2 n + 4} {12 \over{ n + 2}}
{(2n)!\over
{(n!)^{2}} }
= a^{2 n + 4} < {\rm Tr} \varphi^{2n} {\rm
Tr} \varphi^{4} >_{c0}
\end{equation}
The method extends to higher powers of
$m$, but becomes very tedious.
The indirect derivations of ref.[7, 10] are of
course easier.


\newpage
\setcounter{equation}{0}
\renewcommand{\theequation}{B.\arabic
{equation}}
{\bf Appendix B: {Two-point Green's function}}
\vskip 5mm

 The result of two-point correlation function (\ref{N2G2c6})
with non-zero $H_{0}$ and with a Gaussian distribution $P(\varphi)$,
is also derived by considering an ``equation of motion'', which is
obtained by shifting the arbitrary random matrix $\varphi$ to
$\varphi$ + $\epsilon$. The propagator $(1/(z - H_{0} - \varphi)_{ij}
(1/(w - H_{0} - \varphi))_{kl}$ has the corrections of order $\epsilon$
by this shift, and they become
\begin{eqnarray}
& &({1\over{z - H}})_{in}\epsilon_{nm}({1\over{z - H}})_{mj}
({1\over{
w - H}})_{kl}\nonumber\\
&+& ({1\over{z - H}})_{ij}({1\over{w - H}})_{kn}
\epsilon_{nm}({1\over{w - H}})_{ml}\nonumber\\
&-& ({1\over{z - H}})_{ij}({1\over{w - H}})_{kl}(\epsilon_{nm}
\varphi_{mn} + 4 g \epsilon_{nm} (\varphi^{3})_{mn})
\end{eqnarray}

By setting $n = i$, $m = j$ and $k = l$, and summing over these
indices, we obtain the following equation,
\begin{eqnarray}
&&<({1\over{N}}{\rm Tr}{1\over{z - H}})^{2}({1\over{N}}{\rm Tr}
{1\over{w - H}})> + < {1\over{N}}{\rm Tr}{1\over{z - H}}({1
\over{w - H}})^{2} >\nonumber\\
&& - <{1\over{N}}{\rm Tr}({1\over{z - H}}(\varphi + 4 g \varphi^{3}
)){1\over{N}}{\rm Tr}{1\over{w - H}} > = 0
\end{eqnarray}
The second term of above equation is written by
\begin{eqnarray}
< {\rm Tr} ( {1\over{z - H}}({1\over{w - H}})^{2}) > &=&
- {\partial \over{\partial w}}< {\rm Tr}({1\over{z - H}}{1\over{w - H}}
) >\nonumber\\
&=& - {\partial\over{\partial w}}({G(z) - G(w)\over{w - z}})
\end{eqnarray}
By the factorization in the large N limit, the first term of
(B.2) is given by
\begin{eqnarray}
< ( {1\over{N}}{\rm Tr} {1\over{z - H}})^{2} ( {1\over{N}}{\rm
Tr}{1\over{w - H}}) > &=& G^{2}(z)G(w) + 2 G(z) G_{2c}(z,w)
\nonumber\\
&+& G_{2c}(z,z)G(w) + O({1\over{N^{2}}})
\end{eqnarray}
{}From (\ref{NG2}) and (B.2), we have
\begin{eqnarray}
&& 2 G(z) G_{2c}(z,w) + {\partial \over {\partial w}}({G(z) - G(w)
\over{z - w}})\nonumber\\
&& - < {\rm Tr} ({1\over{z - H}}(\varphi + 4 g \varphi^{3}))
{\rm Tr}{1\over{w - H}} >_{c} = 0
\end{eqnarray}

We consider the
two point Green's function
$G_{2c}(z,w)$ based on this equation.
First, we consider the simplest
case: $H_{0} = 0$ and $g = 0$.
In this case, we write the equation for
$G_{2c}(z,w)$,
\begin{equation}
    ( 2 G(z) - z ) G_{2c}(z,w)
= - {\partial\over{\partial w}}({G(z) - G(w)
\over{ z -
w}})
\end{equation}
Since $G(z) = 1/(z - G(z))$ in the case $H_{0} = 0$ and $g = 0$,
we have

\begin{equation}
   {1\over{ 2 G(z) - z}} = {G(z)\over{
G^{2}(z) - 1}} = - {1\over{\sqrt{
z^{2} - 4}}}
\end{equation}
Taking
the derivative of $w$ in (B.6), we have
\begin{eqnarray}
  G_{2c}(z,w) &=&
{1\over{\sqrt{z^{2} - 4}}}{1\over{(z - w)^{2}}}(
      G(z) - G(w) + ( z
- w){G^{2}(w)\over{ 1 - G^{2}(w)}})\nonumber\\
       &=&
{1\over{2(z - w )^{2}}}( {z w - 4 \over{ \sqrt{z^{2} -4}
\sqrt{w^{2} -
4}}} - 1 )
\end{eqnarray}
This coincides with the previous  expression given by
(2.7).

For $ H_{0} \ne 0$ and $ g = 0$ case, we have from
(B.2),

\begin{eqnarray}
 & &  ( 2 G(z) - z ) G_{2c}(z,w) +
{\partial \over{\partial w}}
( {{G(z) - G(w)}\over{ z - w}}
)\nonumber\\
&+& < {\rm Tr}({1\over{z - H_{0} -
\varphi}}H_{0})
{\rm Tr}{1\over{w - H_{0} - \varphi}}>_{c} =
0
\end{eqnarray}
 We diagonalize the matrix $H_{0}$ with the
eigenvalues $\epsilon_{i}$ in the last term.
The diagrams of this
term  are
classified into four different types (Fig.5).
We note that
if we replace $\epsilon_{i}$ by
$(\epsilon_{i} + \Sigma(z) - z) + ( z -
\Sigma(z))$ in the
all four different type diagrams in Fig. 5, we have only an
identity without
any result for $G_{2c}(z,w)$.
Namely, in this case we
have after calculations, ,
\begin{eqnarray}
& &<
{\rm Tr} ( {1\over{ z - H_{0} - \varphi}} H_{0}){\rm Tr} {1\over{ w -
H_{0} - \varphi}}
>_{c} = (z - 2 G(z)) G_{2c}(z,w)\nonumber\\
&+&
{1\over{1 - G(z)\circ G(w)}
} {\partial\over{\partial w}}{\rm Log} ( 1 -
G(z)\circ G(w))
\end{eqnarray}
Inserting this into (B.9), we find that $G_{2c}(z,w)$ cancels out, and
obtain the following identity equation,
\begin{equation}
{\partial\over{\partial w}}({G(z) - G(w)\over{z - w}}) =
    {1\over{1 -
G(z)\circ G(w)}}{\partial \over{\partial w}}
{\rm Log}( 1 - G(z)\circ
G(w))
\end{equation}
To get the expression for  $G_{2c}(z,w)$, we write
separately
$\epsilon_{i}$ for
the diagrams of Fig. 5, (a) and (c) as
$\epsilon_{i} = (\epsilon_{i} + \Sigma(w) - w)
+ ( w - \Sigma(w))$, and
for (b) and (d), $\epsilon_{i} = ( \epsilon_{i}
+ \Sigma(z) - z) + ( z -
\Sigma(z) )$.
Then we have the following expression by summing the
contributions
 of the diagrams (a), (b), (c) and
(d),
\begin{eqnarray}
& &<{\rm Tr}({1\over{z - H_{0} -
\varphi}}H_{0}){\rm Tr} {1\over{w - H_{0} - \varphi}}>_{c}
= ( w -
\Sigma(w) - G(z))G_{2c}(z,w) \nonumber
\\
&+& ( z - w - \Sigma(z) +
\Sigma(w)) G(z)\circ G(z) G_{2c}(z,w)
\nonumber\\
&-& {G^{2}(z)\circ
G(w)\over{(1 - G(z)\circ G(w))(1 - G(w) \circ G(w))}}\nonumber\\
&-&
{(G^{2}(w)\circ G(z))(G(z)\circ G(z))\over{(1 - G(z)\circ G(w))^{2}
( 1
- G(w) \circ G(w))}}
\end{eqnarray}
Using the previous identity of (B.11), we
write the equation for $G_{2c}(z,w)$ by (B.10) and (B.12),
\begin{eqnarray}
& &(G(z) - G(w) - z + w)(1 - G(z)\circ
G(z))G_{2c}(z,w)\nonumber\\
&=& - {1\over{(1 - G(z)\circ G(w))^{2} (
1 - G(w)\circ G(w))}}\nonumber\\
&\times& [G(z)\circ G^{2}(w)( 1-
G(z)\circ G(z))
 - G^{2}(z)\circ G(w) ( 1- G(z)\circ
G(w))]\nonumber\\
& &
\end{eqnarray}
It is easy to see that this equation leads to the previous result (5.3),
which was derived by a diagrammatic analysis$^{4)}$.
 To see this equivalence, we
write the following two
 quantities as
\begin{eqnarray}
G^{2}(z)\circ G^{2}(w) &=& {1\over{N}} \sum_{i} {1\over{(z - \epsilon_{i} -
G(z))^{2}}}
{1\over{(w - \epsilon_{i} - G(w))^{2}}}\nonumber\\
&=&
{G^{2}(z)\circ G(w) - G(z)\circ G^{2}(w)\over{ w - z - G(w) +
G(z)}}
\end{eqnarray}
\begin{equation}
      G^{2}(z)\circ
G(w) = {G(z)\circ G(z) - G(z) \circ G(w)\over
{ w - z - G(w) +
G(z)}}
\end{equation}
and replace these quantities in (B.13), then we
obtain the same result
as (5.3).

 For the non-Gaussian distribution $P(\varphi)$, we obtain the expression
of $G_{2c}(z,w)$, up to order $g$, as (6.9).
When we put $H_{0} = 0$, this expression should be consistent with the
known result, discussed in section 3.
We will show in the following that indeed the expression (6.9) becomes
consistent with the known result when $H_{0} = 0$.

 When $H_{0} = 0$, we have by the
scaling $z
\rightarrow z/a$ and $w \rightarrow
w/a$, from the expression of $G_{2c}(z,w)$ for the Gaussian distribution
$P(\varphi)$,
\begin{eqnarray}
   G_{2c}(z,w) &=& -
{\partial^{2}\over{\partial z \partial w}}
{\rm Log} ( 1 -
G_{0}({z\over{a}}) G_{0}({w\over{a}}) )\nonumber\\
&=&
{\partial^{2}\over{\partial z \partial w}} ({ z - w \over{ z - w
- {z -
{\sqrt{ z^{2} - 4 a^{2}}}\over{2}} + {w - {\sqrt{w^{2} - 4
a^{2}}}
\over{2}}}})
\end{eqnarray}
Now we put $H_{0} = 0$ in the expression of (6.9). Then all circle products
become usual products, and we have also the following equation,
\begin{equation}
   G(z) G(w) = {G(z) - G(w) \over{
w - z - \Sigma(w) + \Sigma(z)}}
\end{equation}
Then we write (6.9) in the following form,
\begin{eqnarray}
    G_{2c}(z,w)& =& -
{\partial^{2}\over{\partial z \partial w}}
{\rm Log} ({ z - w \over{ z - w
+ \Sigma (w) - \Sigma (z) + 4 g G(z)
- 4 g
G(w)}})\nonumber\\
&&
\end{eqnarray}
where the self-energy $\Sigma(z)$ is given as (6.8),
\begin{equation}
    \Sigma(z) = G_{0}(z) - {8 g
G_{0}(z)\over{ 1 - G^{2}_{0}(z)}} -
     { 4 g G^{3}_{0}(z) \over{ 1 -
G^{2}_{0}(z)}} + O(g^{2})
\end{equation}
with

\begin{equation}
    G_{0}(z) = { z - \sqrt{z^{2} - 4}\over
{2}}.
\end{equation}
We expand the quantity $a$ in (B.16) as $a^{2} = 1 - 12 g + O(g^{2})$,
and compare (B.16) and (B.18). Then we find that (6.9) is consistent with
(B.16) when we put $H_{0} = 0$.

\newpage

\begin{center}
{\bf References }
\end{center}
\vskip
3mm

\begin{description}
\item[{1)}] E. Br\'ezin, C. Itzykson, G. Parisi
and J. B. Zuber,
            Comm. Math. Phys. {\bf 59} (1978),
35.
\item[{2)}] E. Br\'ezin and A. Zee, Nucl. Phys. {\bf B40 [FS]} (1993)
613.
\item[{3)}] E. Br\'ezin and A. Zee, C. R. Acad. Sci. {\bf 17} (1993)
735.
\item[{4)}] E. Br\'ezin and A. Zee, Phys. Rev. {\bf E49} (1994)
2588.
\item[{5)}] C. E. Porter, Statistical theories of spectra: fluctuation

(Academic Press, New York,1965)
\item[{6)}] M. L. Mehta, Random
matrices (Academic Press, New York, 1991)
\item[{7)}] A simpler
derivation of this representation of the
            two-point correlation
function was given recently by
            C.W.J. Beenakker, Nucl. Phys.
{\bf B422 } (1994) 515.
\item[{8)}] J. Ambj\o rn and Yu. M. Makeenko,
Modern Physics letters
            {\bf A5} (1990) 1753.
\item[{9)}] F. J.
Dyson, J. Math. Phys. {\bf 3} (1962) 140.
\item[{10)}] The asymptotic
form of orthogonal polynomials with
             a measure\\
${1\over{Z}}\exp[-NV(\lambda)]$, which was used in
             ref[2], has
been recovered through an elegant method due to
             B. Eynard,
Nucl. Phys. {\bf B} (to be published).
\item[{11)}] The electrostatic
problem involved in the calculation
             of the correlation function
was studied recently by B. Jancovici
             and P.J. Forrester (Phys.  Rev. {\bf B} in press), who provided
             thereby an
alternative derivation of (2.2).
\item[{12)}] L. A. Pastur, Theor. Math.
Phys. (USSR) {\bf 10} (1972) 67.
\item[{13)}] E. Br\'ezin, in "2D quantum gravity and random surfaces",
Jerusalem winter
school, edited by D.J.Gross, T.Piran and S. Weinberg, 1992, World Scientific,
Singapore.
\item[{14)}] M. J. Bowick and E. Br\'ezin, Phys. Lett. {\bf B268} (1991) 21.
\item[{15)}] E. Br\'ezin and A. Zee, a preprint NSF-ITP-94-75.
\item[{16)}] C.W.J. Beenakker, Phys. Rev. Lett. {\bf 70} (1993) 1155.
\end{description}

\newpage

{\bf Figure caption}
\vspace{10mm}
\begin{description}
\item[Fig.1]
The self-energy $\Sigma(z)$ is expressed by the product of
             the
vertex part $\Gamma(z)$ and $G(z)$.
\item[Fig.2] The imaginary part of the one-
point Green's function $G(z)$,
for $g=0$, the solid line. The
dotted line represents
the imaginary $G(z)$ for $g=1/2$, and the
broken line represents
             for $g=-1/48$. In all three cases, the
eigenvalues of $H_0$
are $\pm 1$. There is a symmetric counterpart for $z<0$.
\item[Fig.3] The imaginary part of the one-point Green's
function $G(z)$,
for the case in which the eigenvalues of $H_0$ are given uniformly
between $-
$1 and 1.
 The solid
line shows the case of $g=0$, and the dotted
              line corresponds to
$g=1/2,x=11/27$ and the broken line
              shows the case $g=-1/48,
x=4/3$.
\item[Fig.4] The diagrams of order $g$ for $G_{2c}(z,w)$. The broken
             line represents the ladder of $\varphi$ propagators.
\item[Fig.5] Four different diagrams of $< {\rm
Tr} ( {1\over{z - H_{0} - \varphi}}
              H_{0} ) {\rm Tr} {1\over{
w - H_{0} - \varphi}} >$. The
             dotted line means the eigenvalue
$\epsilon_{i}$ of $H_{0}$.
\end{description}
\end{document}